# Metal whisker growth induced by localized, high-intensity DC electric fields

Vamsi Borra[1], Osama Oudat[2], Daniel G. Georgiev[1], Victor G. Karpov[2], Diana Shvydka[3]

1) Department of EECS, University of Toledo, Toledo, Ohio, U.S.A.

2) Department of Physics and Astronomy, University of Toledo, Toledo, Ohio, U.S.A.

3) Department of Radiation Oncology, University of Toledo, Toledo, Ohio, U.S.A.

## ABSTRACT

In this work, a very high, locally applied electric field was used to induce whisker nucleation on an Sn film. The field was generated by using a conductive AFM tip and applying a voltage bias between the sample and the conductive cantilever>The tip-sample separation distance was thus controllable, and any dielectric breakdown could be avoided. At locations where the AFM tip was positioned for an extended period, minuscule whiskers were observed, whose growth direction matched vertical orientation of the field.

## INTRODUCTION

Electrically conductive hair-like structures, referred to as whiskers, can bridge the gap between densely spaced electronic devices and components. This can lead to current leakage and short circuits, causing significant losses and, in some cases, catastrophic failure in various systems, including such in the automotive, aerospace and other industry[1], [2]. None of the whisker growth models proposed to date is capable of answering consistently and universally why whiskers grow, in the first place. Although there are several generally accepted factors (intermetallic compound formation, tensile / compressive stress, Sn oxide layer presence, coefficient of thermal expansion (CTE) mismatch that influence the whisker growth), the most important factor in whiskering is still a matter of dispute. A recent theory[3]–[8], which considers the imperfections (small patches of net positive or negative electric charges) on metal surface, details the quantitative estimates of the metal whisker nucleation along with their growth rates and length distributions. According to this theory, the anomalous electric field (E) formed due to the imperfections will govern the whisker development in those areas. In addition, an external electric field, which can be either constant (DC) or varying with time (AC) at high frequencies (including optical), can also contribute to nucleation and promote their growth by means of lowering the free energy of the system.

Here, we present a new way of controlling the growth of whiskers by using localized high-intensity DC electric fields by using an AFM setup and we refer to it as whisker engineering. A current sensing AFM scanner with a conducting cantilever was utilized. The electric field was generated by applying a voltage bias between the sample and the conductive cantilever, which is maintained at a known distance, without causing any dielectric breakdown. SEM examination of samples at the points where the AFM tip was positioned for an extended period was performed before and after electric field application. Minuscule whiskers were observed; whose growth direction matched the direction of the field. The observed whisker formation can be used to design a new non-destructive readily implementable accelerated failure testing procedure as well as in other applications.

**EXPERIMENTAL DETAILS**

In this study, optically smooth Sn and Zn thin, films fabricated by vacuum evaporation[9], are subjected to high-intensity DC electric field to instigate the whisker growth. Here, we took advantage of controlling the AFM's tip precisely from the sample's surface. A positively biased conductive AFM tip was maintained at a known distance from the sample, which is negatively biased. The electric fields resulted from this setup was confined to a very small area of the sample.

For this experiment, the PicoPlus AFM setup was modified to use the ultra-sharp AFM cantilever, coated with conductive platinum film, in Current Sensing AFM (CSAFM) mode. CSAFM requires a special pre-amplifier and cone that holds the conductive tip. The bias voltage is applied to the sample while maintaining the tip at virtual ground. Here, the generated current, when in contact with sample, assists in constructing a conductivity image. However, we have not used this imaging mode in our case, as we are interested in generating a localized E-field rather than a surface topography. The conductive tip (probe) was first clipped to the CSAFM nose using a retracting spring on to the retaining guides. Then the nose assembly, along with the tip, was inserted into the scanner. After connecting the required cables, the laser spot on the cantilever and the detector position were aligned (center of four quadrants). The sample was placed on the sample plate and the two adjacent pogo pins were used to secure it. Special care was taken to electrically isolate the sample from the sample plate. The platinum electrode was placed on the sample and then connected to the spring-loaded electrode clip by lifting it. The continuity was tested using digital multimeter to ensure the steady connection, then the sample plate was placed under microscope. Simultaneously, the video system and its illuminator were adjusted such that the tip and sample are in focus. This video system was employed to aligning the laser spot on the AFM tip and observe the AFM tip's position with respect to the reference point. Several reference points/patterns are marked on the samples before exposing the sample to the AFM's electric field. One such sample with reference points marked by using a laser is shown in the insert of Fig 1. The samples were observed under SEM before performing the AFM experiment, to detect any whiskers.

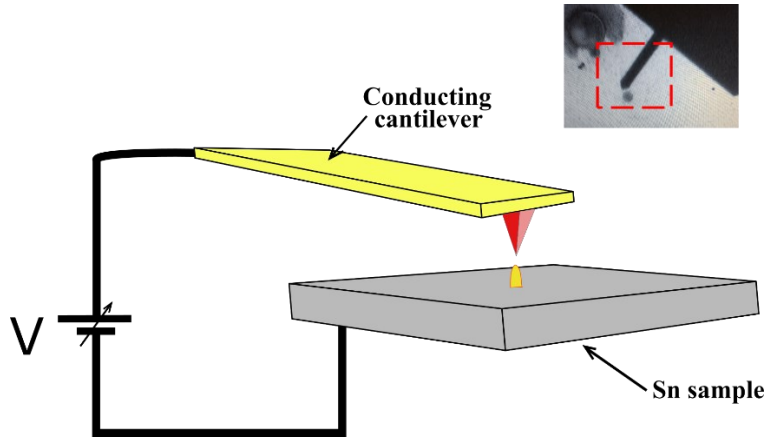

**Figure 1. Experimental setup, insert on the top right shows the AFM tip positioned at a reference point.**

Tip lowering sequence was performed, using a high precision stepper motor from PicoPlus software, until tip touches the sample surface. Then, the tip was withdrawn vertically from the sample surface to a distance of ~15µm through software. A voltage bias of 10V was applied to the sample (through Pt electrode) using the potentiostat window within PicoPlus software, and was left for several hours (~ 22-24), see Fig 1 for illustration. Special attention was given to the applied bias voltage and the sample-tip separation distance such that the generated field does not exceed the dielectric strength of the air. This process, of applying bias, was repeated on several spots on a same sample. Sample was then scanned under SEM to identify the whisker nucleation sites.

**DISCUSSION**

The approximate generated field (V/m) can be calculated by using $E = \frac{V}{d}$. Here, $V$ is the applied bias and $d$ is the distance between the tip and the sample. The calculated E-field from this setup is about ~$6.7 \times 10^5 V/m$, which is an order of magnitude lower than the dielectric breakdown of air.

Upon examining the areas near the reference points, where the AFM tip was positioned, we have observed single miniscule whiskers at the vicinity of that point, as shown in Fig. 2(a) and (b). Out of a total of eleven locations treated with this AFM-generated field, five locations have showed a sign of whisker nucleation. Whereas the areas surrounding the spot have not shown any whisker nucleation, as seen in Fig. 4. These surrounding areas were examined carefully and thoroughly in multiple SEM image such as the one shown in Fig. 4b.

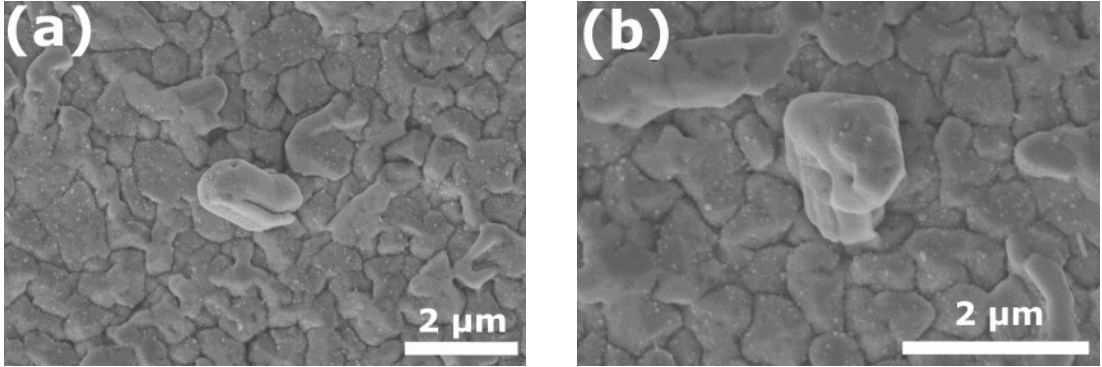
**Figure 2: Whiskers observed at the spots where the AFM tip was placed.**

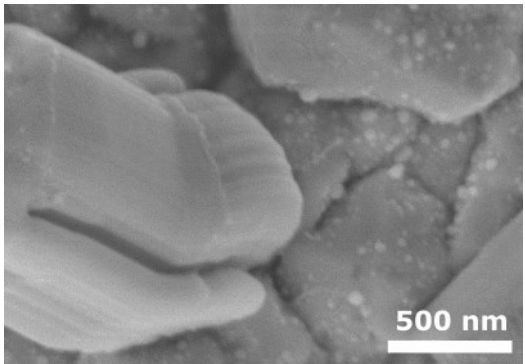
**Figure 3: High-magnification SEM of 2(a) confirming the whisker.**

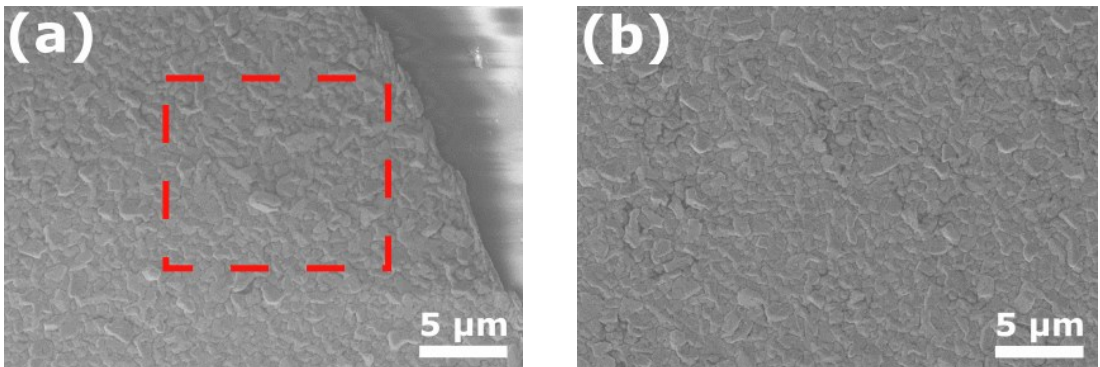
**Figure 4: (a) Low-magnification SEM scan of whisker in Fig 2(a), (b) Surface away from the irradiated spot.**

The characteristic ridges on the whiskers can be observed on the high-magnification SEM scan, Fig. 3, of the whisker in Fig. 2(a). These ridges, which can also be seen in Fig 2(b), confirm the whisker formation rather than a hillock or any other structure. A low-magnification SEM scan of the same location, Fig 2(a), is shown in Fig 4(a). In this image, no other whisker or similar nodules can be observed in the vicinity of the identified whisker (highlighted in red box). A similar surface scan away from the irradiated spot is shown in Fig 4(b), where no whiskers are detected as well.

The growth direction of both whiskers (Fig 2(a) and (b)) are almost in the vertical direction, which align with the applied E-field. This growth observations can be attributed to the mechanism described in electrostatic theory[3], which has forecasted the whisker development.

The estimated field strengths resulted from this experiment are in the order of ~$6.7 \times 10^5 V/m$, which are on the lower side of the theoretical predictions and E-field strengths observed in other experiments [6], [8], [10]. E-field enhancement [11] at the tip of the AFM probe, which falls in the frame of the electrostatic theory[3] of whiskers, is a possible explanation for the observed whisker growth in this experiment. One plausible explanation for the observed whisker growth, still with less intense fields, can be due to the high E-field intensity at the tip of the AFM. As far as the relatively low yield (5 out of 11 experimental points showed whisker growth), this can be attributed to the deficiency of the required E-field or surface irregularities/contamination on the film, which are not identified at this point.

## CONCLUSIONS

A very high local electric field, generated using a conductive AFM tip, resulted in direct whisker nucleation and some growth. The field was obtained by applying a voltage between the sample and the conductive cantilever and maintaining a known distance. SEM examination of the sample was performed before and after electric field application. Minuscule whiskers were observed, whose growth direction matched with the orientation of the field, at the points where the AFM tip was positioned for an extended period of time. The observed whisker growth can be attributed to presence of high electric fields as well as some tip-enhancement near the tip of the AFM probe. This experimental observation may have practical value in the development of non-destructive accelerated failure field-based testing methods for whiskering of electronic components, which is currently not available due to the erratic nature of whisker growth.

## ACKNOWLEDGMENTS

The authors would like to thank the University of Toledo's EECS department for providing the financial support to present this work. This work has benefitted from technical help with the deposition of films, provided by the Wright Center for Photovoltaics Innovation and Commercialization (PVIC) at the University of Toledo.